\documentclass[a4paper,11pt]{article}
\pdfoutput=1 

\usepackage{jheppub} 

\usepackage[normalem]{ulem}
\usepackage{amsmath,amssymb, mathrsfs, float}
\usepackage{epsfig}
\usepackage{graphicx}                   
\usepackage{url}
\usepackage{color}
\usepackage{slashed}
\usepackage{multirow}
\usepackage{placeins}
\usepackage[dvipsnames]{xcolor}
\usepackage{epstopdf}
\usepackage{soul}
\usepackage{tikz}
\usepackage[capitalise,english]{cleveref}
\usepackage{siunitx}
\usepackage{xspace}
\usepackage{booktabs}
\usepackage{tabularx}
\usepackage{gensymb}
\usetikzlibrary{trees}
\usetikzlibrary{decorations.pathmorphing}
\usetikzlibrary{decorations.markings}

\usepackage[compat=1.1.0]{tikz-feynman}
\usepackage{subcaption}

\newcommand*\cancel[2][thin]{
    \tikz[baseline] \node [strike out,draw,anchor=text,inner sep=0pt,text=black,#1]{#2};
}
\usetikzlibrary{arrows, shapes, patterns.meta}
\usepackage{pifont}

\newcommand{\gap}[2]{\hspace{#1}#2\hspace{#1}}

\allowdisplaybreaks[4]

\title{\boldmath Cogenesis of visible and dark matter in a scotogenic model}

\author[a]{Debajit Bose,}
\author[a]{Rohan Pramanick,}
\author[a]{and Tirtha Sankar Ray}

\affiliation[a]{Department of Physics, Indian Institute of Technology Kharagpur, Kharagpur 721302, India}

\emailAdd{debajitbose550@gmail.com}
\emailAdd{rohanpramanick25@gmail.com}
\emailAdd{tirthasankar.ray@gmail.com}

\abstract{Within a  scotogenic neutrino mass model we explore the cogenesis of matter from  the  CP violating decay of  a heavy $\mathbb{Z}_2$-odd right handed neutrino that simultaneously populates the visible and a multipartite dark sector. The relic density of a sub-GeV scale freeze-in dark matter is generated by the late time decay of the next-to-lightest dark particle dynamically regulated by an interplay with the thermal scattering processes. We show that this model can simultaneously explain visible matter asymmetry and provide a cosmologically viable sub-GeV dark matter while remaining in consonance with the neutrino parameters and flavour observables.}

\begin{document} 

\newcommand{\eq}{\text{eq}}
\newcommand{\yNtwo}{y_{N_2}}
\newcommand{\yNone}{y_{N_1}}
\newcommand{\ySE}{y_{\Sigma \eta}}
\newcommand{\ydE}{y_{\delta \eta}}
\newcommand{\ydL}{y_{\delta L}}

\newcommand{\gamNtwo}{\gamma_{N_2}}
\newcommand{\gamE}{\gamma_{\eta}}
\newcommand{\gamNtwoNtwoTOLLb}{\gamma^{N_2 N_2}_{L \overline{L}}}
\newcommand{\gamNtwoNoneTOLLb}{\gamma^{N_2 N_1}_{L \overline{L}}}
\newcommand{\gamNtwoNtwoTOEEb}{\gamma^{N_2 N_2}_{\eta \overline{\eta}}}
\newcommand{\gamNtwoNoneTOEEb}{\gamma^{N_2 N_1}_{\eta \overline{\eta}}}
\newcommand{\gamNtwoLTONoneL}{\gamma^{N_2 L}_{N_1 L}}
\newcommand{\gamNtwoETONoneE}{\gamma^{N_2 \eta}_{N_1 \eta}}
\newcommand{\gamEETOLbLb}{\gamma^{\eta \eta}_{\overline{L}\, \overline{L}}}
\newcommand{\gamEEbTOHHb}{\gamma^{\eta \overline{\eta}}_{H \overline{H}}}
\newcommand{\gamEETOHH}{\gamma^{\eta \eta}_{H H}}
\newcommand{\gamLETOLbEb}{\gamma^{L \eta}_{\overline{L} \overline{\eta}}}

\maketitle
\flushbottom

\newcommand{\tabFieldContent}{
\begin{table*}[t]
    \centering
    \renewcommand{\arraystretch}{1.25}
    \begin{tabular}{|c|c|c|c|c|}
		\hline
        & $SU(3)_c$ & $SU(2)_L$ & $U(1)_Y$ & $\mathbb{Z}_2$ \tabularnewline
        \hline
        \hline
		$L_L$ & 1 & 2 & $-\frac{1}{2}$ & $+$ \tabularnewline \hline
		$H$   & 1 & 2 & $ \frac{1}{2}$ & $+$ \tabularnewline \hline
		$e_R$ & 1 & 1 & $-1$            & $+$ \tabularnewline \hline
		$\eta$& 1 & 2 & $ \frac{1}{2}$ & $-$ \tabularnewline \hline
		$N_i$ & 1 & 1 & $ 0 $ 			& $-$ \tabularnewline \hline
    \end{tabular}
	\renewcommand{\arraystretch}{1.0}
	\caption{Charge assignment and the field content of the Scotogenic model.}
	\label{tab:Field_Content}
\end{table*}
}

\newcommand{\tabParamRange}{
\begin{table*}[!t]
    \centering
    \renewcommand*{\arraystretch}{1.25}
    \resizebox{1\linewidth}{!}{
    \begin{tabular}{|c||c|c|c|c|c|}
        \hline
        Parameter & $\mathrm{Re}[Y_{\alpha 1}], \ \mathrm{Im}[Y_{\alpha 1}]$ & $\mathrm{Re}[Y_{\alpha 2}], \ \mathrm{Im}[Y_{\alpha 2}]$ & $\mathrm{Re}[Y_{\alpha 3}], \ \mathrm{Im}[Y_{\alpha 3}]$ & $\lambda_3$ & $\lambda_5$ \tabularnewline
        \hline
        Range & $\left[ 5 \times 10^{-9}, \ 5 \times 10^{-7} \right]$ & [0.01, \ 0.1] & [0.01, \ 0.5] & [1.0, \ 4$\pi$] & [0.01, \ 0.1] \tabularnewline
        \hline
    \end{tabular}
    }
    \renewcommand*{\arraystretch}{1.25}
    \caption{The range of various parameters implemented in the numerical scan, where $\alpha = 1, 2$ and 3.}
    \label{tab:range_free_param}
\end{table*}
}

\newcommand{\tabBPdetails}{
\begin{table*}[t]
    \centering
    \renewcommand{\arraystretch}{1.5}
    \resizebox{1\textwidth}{!}{
    \begin{tabular}{|c|c|c|c|c|c|}
\hline
\gap{0.25cm}{$M_{N_3}$ [GeV]} & \gap{0.25cm}{$M_{N_2}$ [GeV]} & \gap{0.25cm}{$M_\eta$ [GeV]} & \gap{0.85cm}{$\lambda_3$} & \gap{0.85cm}{$\lambda_4$} & $\lambda_5$\tabularnewline
\hline
\hline
$5 \times 10^{10}$ & $5 \times 10^{9}$ & $5 \times 10^{8}$ & 3.276 & 0 & $3.096 \times 10^{-2}$ \tabularnewline
\hline
\hline
\multicolumn{6}{|c|}{Yukawa couplings ($Y_{\alpha i}$)}\tabularnewline
\hline
\hline
\multicolumn{6}{|c|}{$\begingroup \setlength\arraycolsep{8pt} \begin{pmatrix}
(5.410 + i \ 5.421) \times 10^{-8} & (2.280 + i \ 1.108) \times 10^{-2} & (1.009 + i \ 1.294) \times 10^{-2} \\[-0.5ex] 		(7.210 + i \ 4.361) \times 10^{-8} & (1.408 + i \ 4.132) \times 10^{-2} & (1.496 + i \ 0.241) \times 10^{-1} \\[-0.5ex] 		(8.612 + i \ 3.060) \times 10^{-8} & (1.758 + i \ 1.203) \times 10^{-2} & (0.292 + i \ 1.571) \times 10^{-1}          \end{pmatrix} \endgroup$}\tabularnewline
\hline
\end{tabular}
    }
    \renewcommand{\arraystretch}{1.0}
    \caption{Details of the benchmark point for the cogenesis scenario.}
    \label{tab:bp_details}
\end{table*}
}

\newcommand{\tabBPdetailsOld}{
\begin{table*}[h]
    \centering
    \renewcommand{\arraystretch}{1.5}
    \resizebox{1\textwidth}{!}{
    \begin{tabular}{|c|c|c|c|c|c|}
\hline
\multicolumn{6}{|c|}{Yukawa couplings ($Y_{\alpha i}$)}\tabularnewline
\hline
\hline
\multicolumn{6}{|c|}{$\begingroup \setlength\arraycolsep{8pt} \begin{pmatrix}             (4.170 + i \ 3.351) \times 10^{-8} & (1.605 + i \ 2.334) \times 10^{-2} & (1.050 + i \ 1.308) \times 10^{-2} \\[-0.5ex] 			(9.255 + i \ 3.076) \times 10^{-8} & (1.109 + i \ 3.529) \times 10^{-2} & (1.705 + i \ 0.394) \times 10^{-1} \\[-0.5ex] 			(4.742 + i \ 1.239) \times 10^{-8} & (2.824 + i \ 1.228) \times 10^{-2} & (0.382 + i \ 1.275) \times 10^{-1}          \end{pmatrix} \endgroup$}\tabularnewline
\hline
\end{tabular}
    }
    \renewcommand{\arraystretch}{1.0}
    \caption{Details of the benchmark point for the cogenesis scenario with higher value of $\lambda_3$.}
    \label{tab:bp_details_old}
\end{table*}
}

\newcommand{\tabBPoutput}{
\begin{table*}[t]
    \centering
    \renewcommand{\arraystretch}{1.5}
    \begin{tabular}{|c|c|c|}
        \hline
        \multicolumn{3}{|c|}{Neutrino oscillation observables within $3\sigma$ in normal hierarchy}\tabularnewline
        \hline \hline
        \gap{0.75cm}{$\Delta m_{12}^2 \left[\text{eV}^2 \right]$} & \gap{0.75cm}{$\Delta m_{13}^2 \left[\text{eV}^2 \right]$} & $\sum m_\nu$ [eV] \tabularnewline
        \hline
        $7.285 \times 10^{-5}$ & $2.562 \times 10^{-3}$ & $0.059$ \tabularnewline
        \hline
        $\theta_{12}$ & $\theta_{23}$ & $\theta_{13}$ \tabularnewline
        \hline
        $31.333\degree$ & $42.279\degree$ & $8.343\degree$ \tabularnewline
        \hline
        $\delta_{\text{CP}}$ & $J_{\text{CP}}$ & $|m_{\beta \beta}|$ [eV] \tabularnewline
        \hline
        $207.870\degree$ & $-0.031$ & $2.980 \times 10^{-3}$ \tabularnewline
        \hline
        \hline
        \multicolumn{3}{|c|}{Quantities related to cogenesis}\tabularnewline
        \hline \hline
        $\epsilon_2$ & $Y_{\delta B}$ & $M_{N_1}$ satisfying $\Omega_{\text{DM}}h^2 = 0.12$ \tabularnewline
        \hline
        $7.640 \times 10^{-5}$ & $8.844 \times 10^{-11}$ & $3.827$ MeV \tabularnewline
        \hline \hline
        \multicolumn{3}{|c|}{Quantities related to lepton flavour violating processes}\tabularnewline
        \hline \hline
        $\text{Br}(\mu \rightarrow e + \gamma)$ & $\text{Br}(\mu \rightarrow 3e)$ & electron dipole moment $|d_e|/e$ \tabularnewline
        \hline
        $1.157 \times 10^{-41}$ & $3.380 \times 10^{-50}$ & $1.572 \times 10^{-55}$ cm \tabularnewline
        \hline
    \end{tabular}
\caption{Relevant quantities for the benchmark point satisfying neutrino oscillation data, baryon asymmetry, DM relic abundance and all other constraints from sum over neutrino masses, structure formation, LFV processes, EDM measurements and perturbative unitarity.}
\label{tab:bp_details_output}
\end{table*}
}

\newcommand{\tabBPoutputOld}{
\begin{table*}[t]
    \centering
    \renewcommand{\arraystretch}{1.5}
    \begin{tabular}{|c|c|c|}
        \hline
        \multicolumn{3}{|c|}{Neutrino oscillation observables within $3\sigma$ in normal hierarchy}\tabularnewline
        \hline \hline
        \gap{0.75cm}{$\Delta m_{12}^2 \left[\text{eV}^2 \right]$} & \gap{0.75cm}{$\Delta m_{13}^2 \left[\text{eV}^2 \right]$} & $\sum m_\nu$ [eV] \tabularnewline
        \hline
        $7.257 \times 10^{-5}$ & $2.567 \times 10^{-3}$ & $0.059$ \tabularnewline
        \hline
        $\theta_{12}$ & $\theta_{23}$ & $\theta_{13}$ \tabularnewline
        \hline
        $33.535\degree$ & $50.314\degree$ & $8.918\degree$ \tabularnewline
        \hline
        $\delta_{\text{CP}}$ & $J_{\text{CP}}$ & $|m_{\beta \beta}|$ [eV] \tabularnewline
        \hline
        $186.137\degree$ & $-0.033$ & $3.722 \times 10^{-3}$ \tabularnewline
        \hline
        \hline
        \multicolumn{3}{|c|}{Quantities related to cogenesis}\tabularnewline
        \hline \hline
        $\epsilon_2$ & $Y_{\delta B}$ & $M_{N_1}$ satisfying $\Omega_{\text{DM}}h^2 = 0.12$ \tabularnewline
        \hline
        $6.116 \times 10^{-5}$ & $8.779 \times 10^{-11}$ & $9.581$ MeV \tabularnewline
        \hline \hline
        \multicolumn{3}{|c|}{Quantities related to lepton flavour violating processes}\tabularnewline
        \hline \hline
        $\text{Br}(\mu \rightarrow e + \gamma)$ & $\text{Br}(\mu \rightarrow 3e)$ & electron dipole moment $|d_e|/e$ \tabularnewline
        \hline
        $1.072 \times 10^{-41}$ & $3.160 \times 10^{-50}$ & $1.439 \times 10^{-55}$ cm \tabularnewline
        \hline
    \end{tabular}
\caption{Relevant quantities for the benchmark point satisfying neutrino oscillation data, baryon asymmetry, DM relic abundance and all other constraints from sum over neutrino masses, structure formation, LFV processes, EDM measurements and perturbative unitarity.}
\label{tab:bp_details_output}
\end{table*}
}

\newcommand{\figParamSpaceLeptoDM}{
\begin{figure*}[t]
    \centering
    \begin{subfigure}{0.495\linewidth}
        \centering
        \includegraphics[width=1\textwidth]{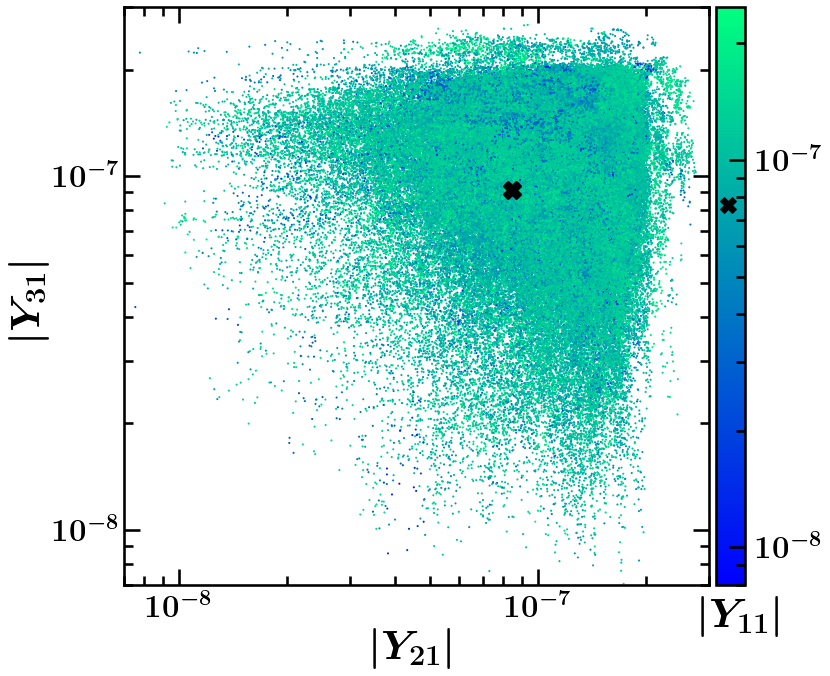}
        \caption{}
        \label{fig_lepto_only_N1}
    \end{subfigure}
    \begin{subfigure}{0.495\linewidth}
        \centering
        \includegraphics[width=1\textwidth]{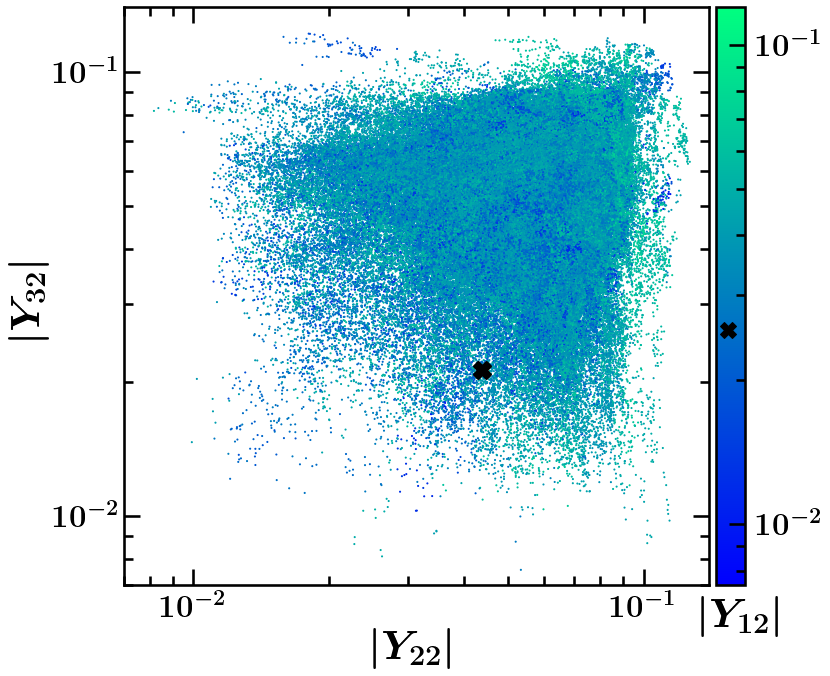}
        \caption{}
        \label{fig_lepto_only_N2}
    \end{subfigure}
    \begin{subfigure}{0.495\linewidth}
        \centering
        \includegraphics[width=1\textwidth]{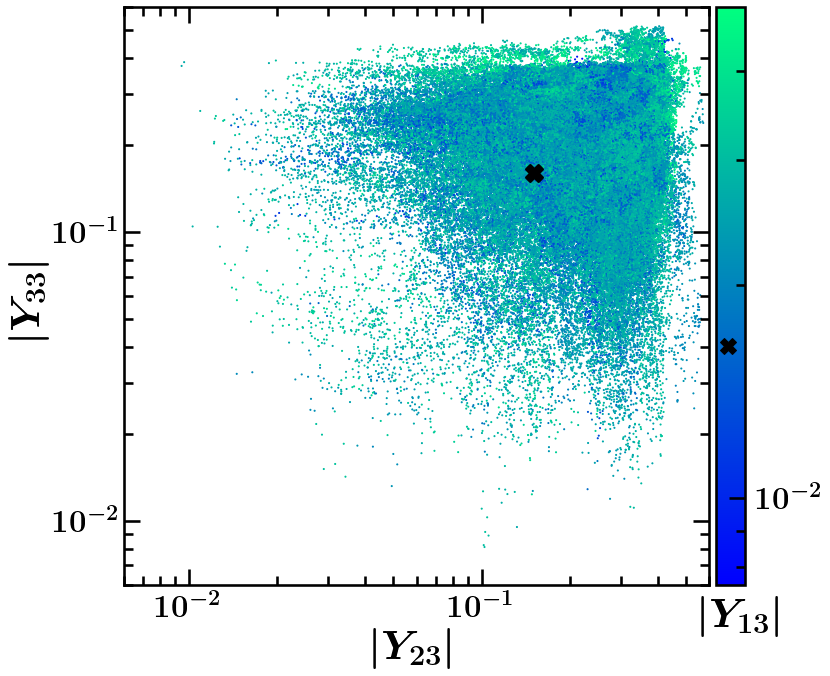}
        \caption{}
        \label{fig_lepto_only_N3}
    \end{subfigure}
    \begin{subfigure}{0.495\linewidth}
        \centering
        \includegraphics[width=1\textwidth]{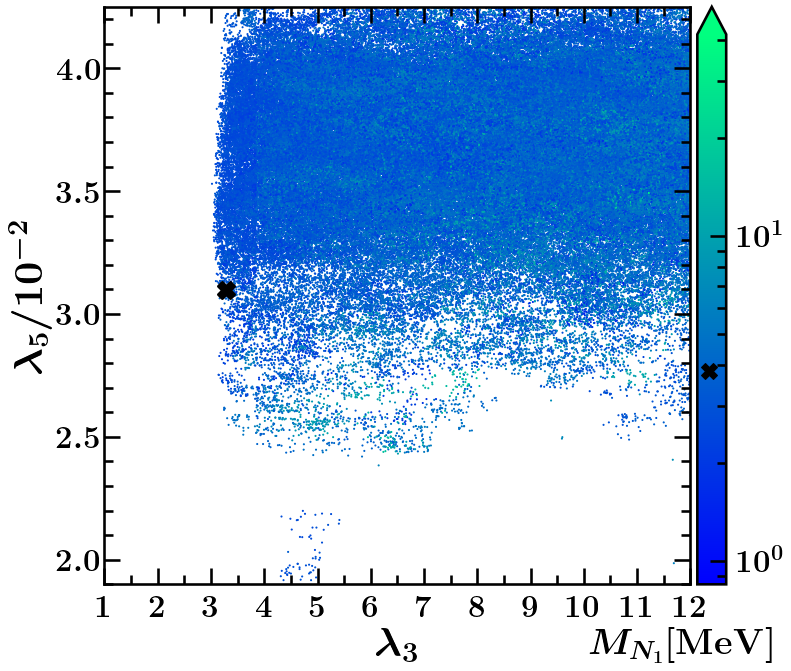}
        \caption{}
        \label{fig_lepto_only_quartics}
    \end{subfigure}
    \caption{Parameter space yielding correct BAU and relic abundance while satisfying the structure formation bounds as presented in terms of the Yukawa couplings of \eqref{fig_lepto_only_N1} $N_1$, \eqref{fig_lepto_only_N2} $N_2$, \eqref{fig_lepto_only_N3} $N_3$ RHN and \eqref{fig_lepto_only_quartics} quartic couplings between Higgs and the inert scalar doublet. The benchmark point in Table III is indicated by \ding{54}.}
    \label{fig_lepto_only}
\end{figure*}
}

\newcommand{\figBPforReacDen}{
\begin{figure*}[t]
    \centering
    \begin{subfigure}{0.495\linewidth}
        \centering
        \includegraphics[width=1\textwidth]{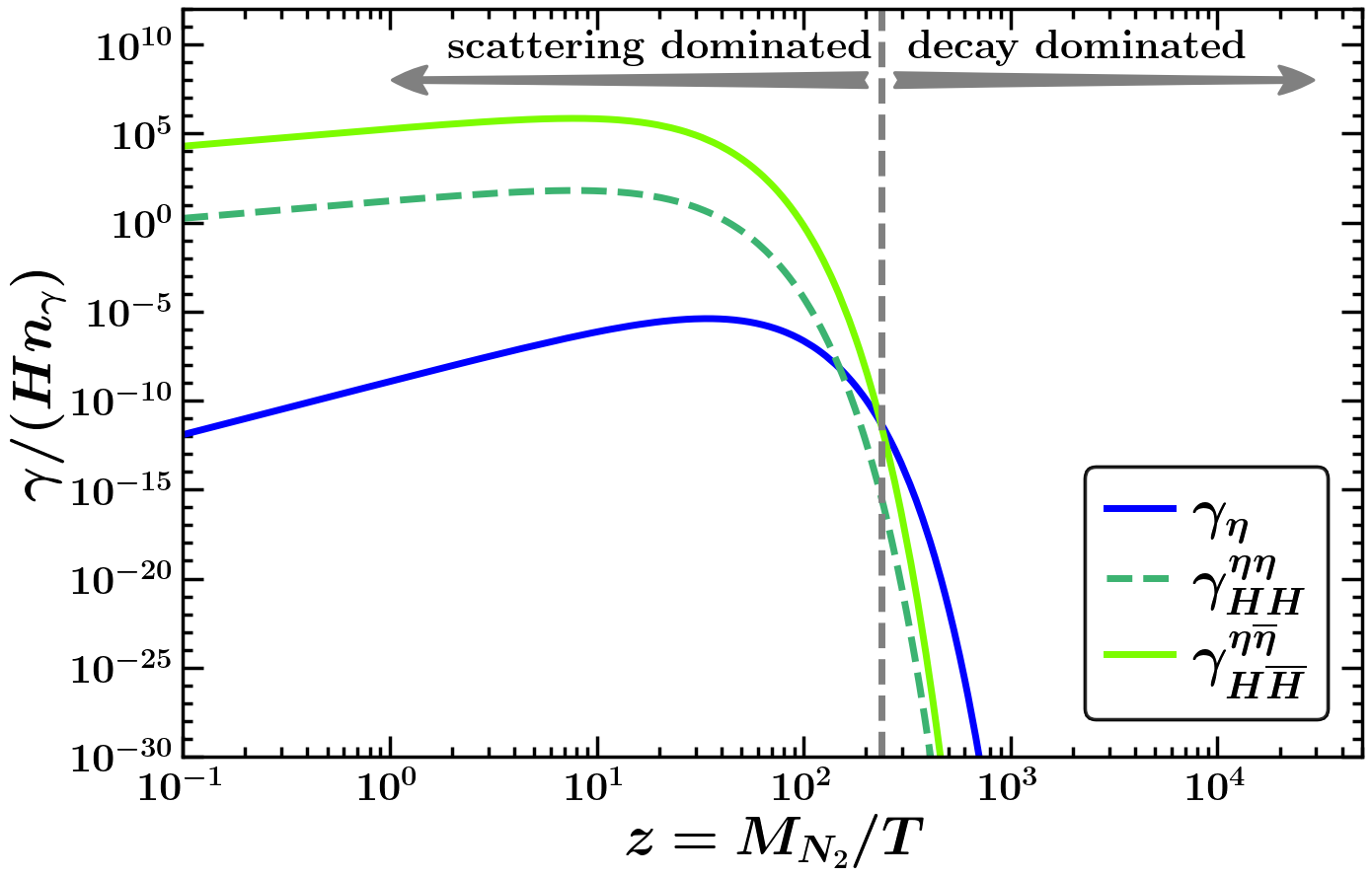}
        \caption{}
        \label{fig:reac_den}
    \end{subfigure}
    \begin{subfigure}{0.495\linewidth}
        \centering
        \includegraphics[width=1\textwidth]{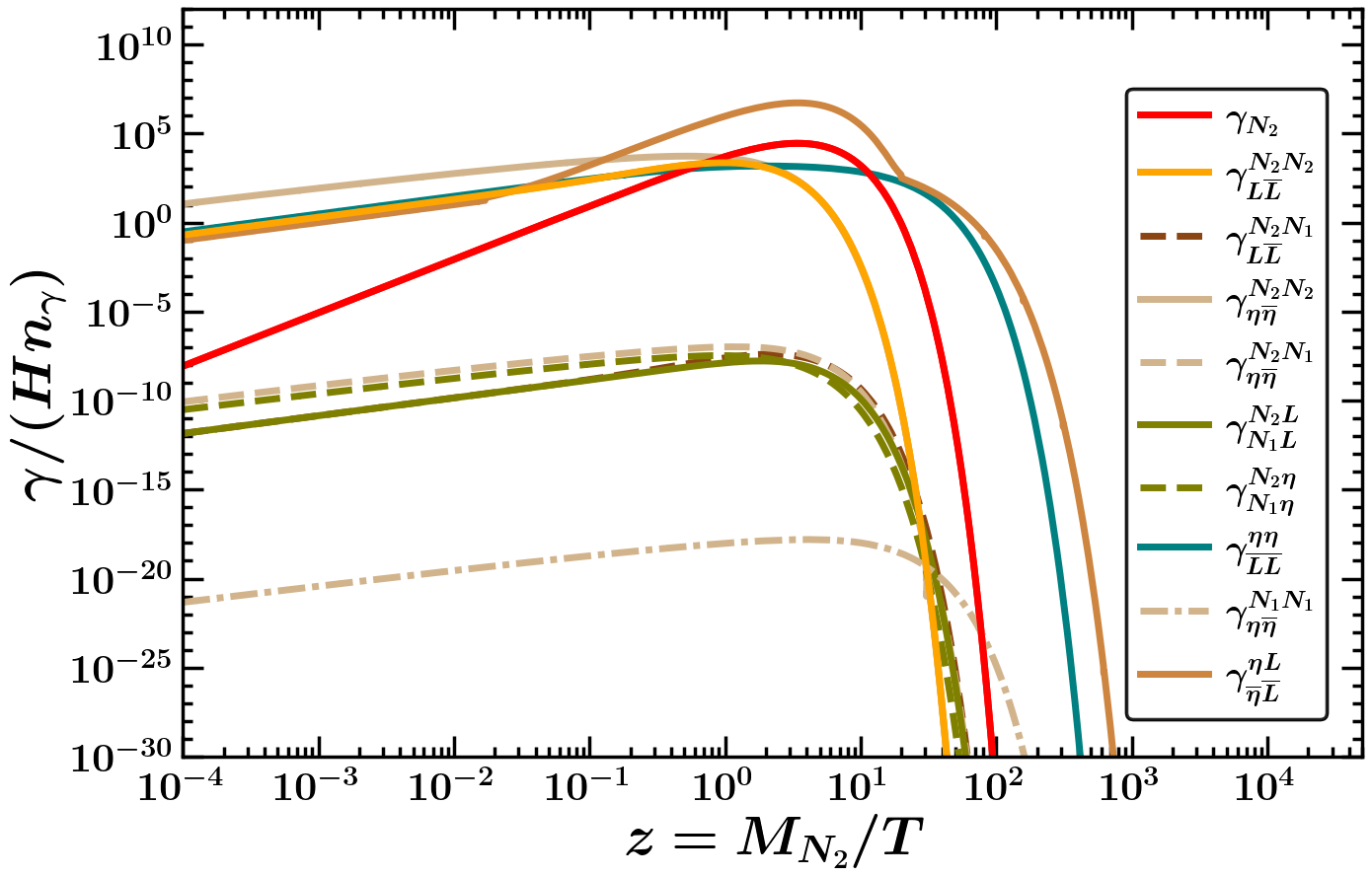}
        \caption{}
        \label{fig:reac_den_all}
    \end{subfigure}
    \caption{Left panel \ref{fig:reac_den} shows the decay and scattering reaction densities of the inert scalar doublet with respect to $z$. The vertical dashed line represents the equality of both these reaction densities. All other reaction densities utilized in Eq. \ref{eq:boltz_full} are shown in the right panel \ref{fig:reac_den_all} for the benchmark point given in Table \ref{tab:bp_details}.}
\end{figure*}
}

\newcommand{\figBPforY}{
\begin{figure*}[t]
    \centering
    \includegraphics[width=0.65\textwidth]{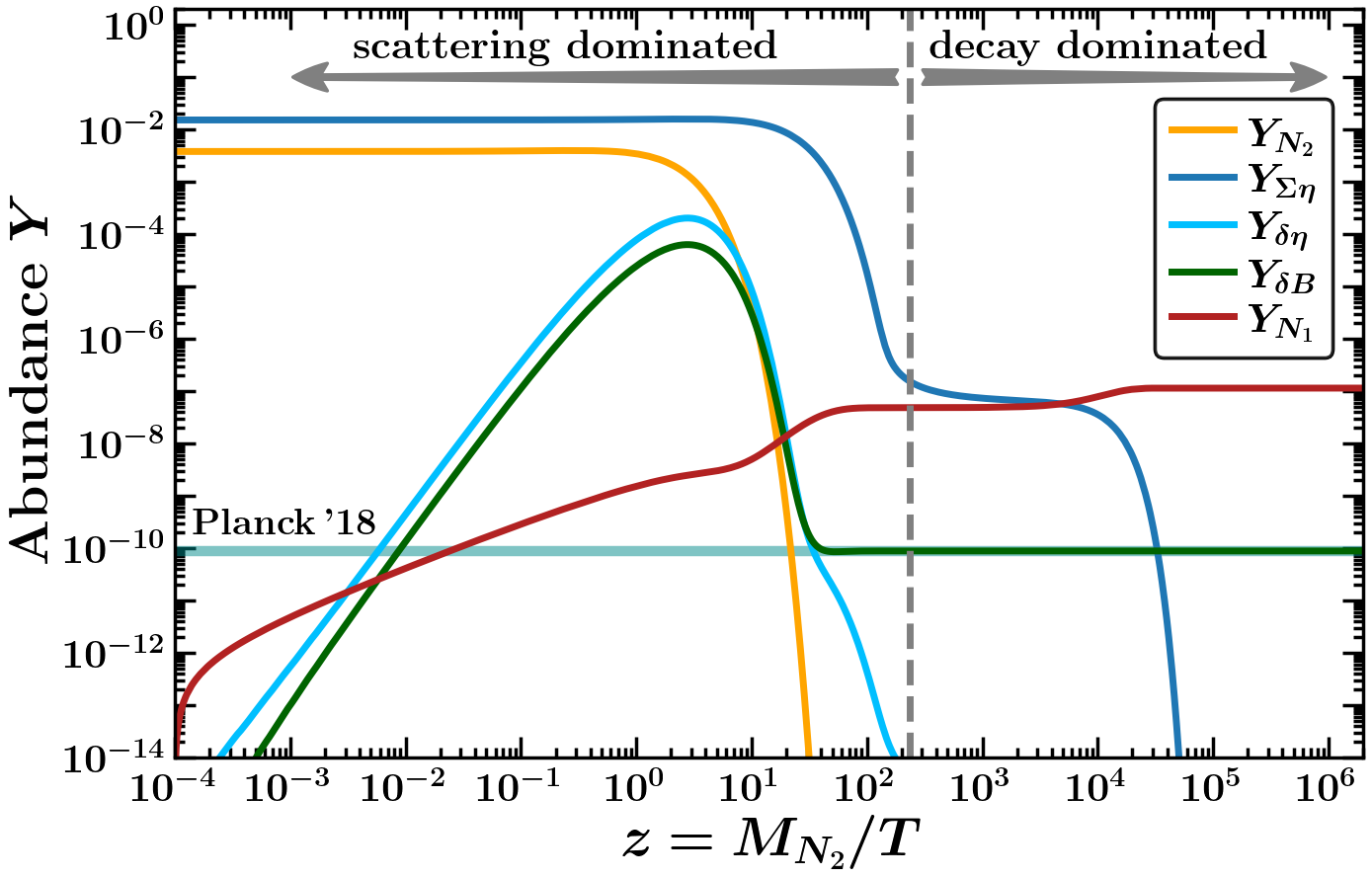}
    \caption{Evolution of abundances for various species involved in cogenesis for the benchmark point in Table \ref{tab:bp_details}.}
    \label{fig:Y}
\end{figure*}
}

\newcommand{\figNeutrinoMassFeynmanDiagram}{
\begin{figure*}[t]
\centering \scalebox{1.0}{
\begin{tikzpicture}[baseline=(current bounding box.center), dot/.style={draw,circle,minimum size=1.75mm,inner sep=0pt,outer sep=0pt,fill=black}]
	\begin{feynman}
		\vertex (a1) ;
		\vertex [right = 1.5cm of a1] (a2) ;
		\vertex [right = 2cm of a2] (a3) ;
		\vertex [right = 2cm of a3] (a4) ;
		\vertex [right = 1.5cm of a4] (a5) ;
		\node   [above = 2cm of a3, dot] (b1) ;
		\vertex [above = 0.125cm of a3] (c1) {\large $N_{1, 2, 3}$} ;
		\vertex [left = 2.75cm of c1] (c2) {\large $L$} ;
		\vertex [right = 2.75cm of c1] (c3) {\large $\overline{L}$} ;
		\vertex [above left = 1cm and 2cm of b1] (b1_leftup) {$\langle H \rangle$};
		\vertex [above right = 1cm and 2cm of b1] (b1_rightup) {$\langle H \rangle$};

		\diagram{
		(a1) -- [fermion, very thick] (a2);
		(a2) -- [plain, very thick, insertion = 0.5] (a4);
		(a4) -- [anti fermion, very thick] (a5);
		(a2) -- [scalar, quarter left, very thick, edge label = {\large $\eta$}] (b1);
		(b1) -- [scalar, quarter left, very thick, edge label = {\large $\eta$}] (a4);
		(b1) -- [scalar, very thick] (b1_leftup);
		(b1) -- [scalar, very thick] (b1_rightup);
		};
	\end{feynman}
\end{tikzpicture}
}
\caption{Neutrino mass generation in the Scotogenic model}
\label{fig:Neutrino_Mass_Feynman_Diagram}
\end{figure*}
}

\newcommand{\figMDMvsSummnu}{
\begin{figure}
    \centering
    \includegraphics[width = 0.45\textwidth]{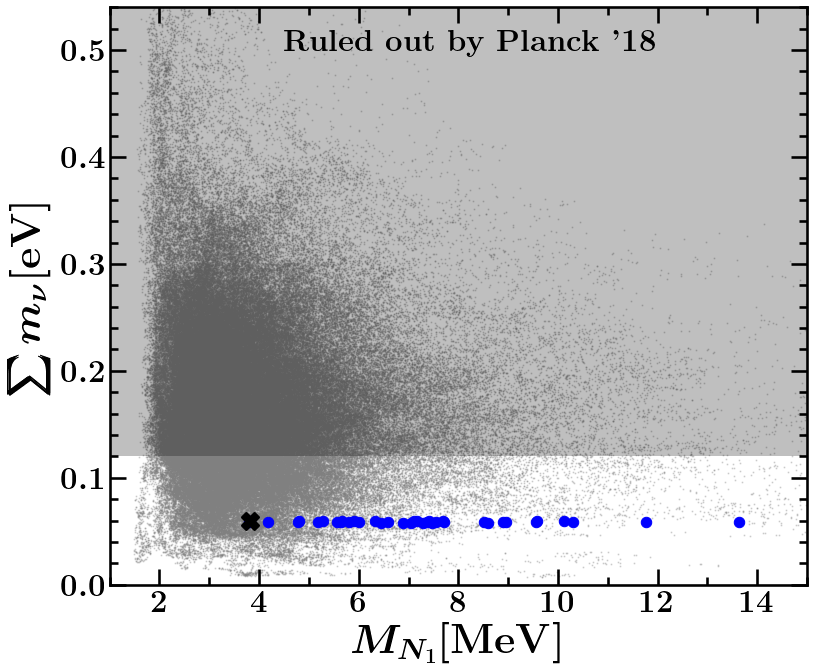}
    \caption{Parameter space consistent with the upper bound on sum over neutrino masses.}
    \label{fig_MN1_vs_summnu}
\end{figure}
}

\newcommand{\figParamSpaceAll}{
\begin{figure}
    \centering
    \begin{subfigure}{0.495\linewidth}
        \centering
        \includegraphics[width=1\textwidth]{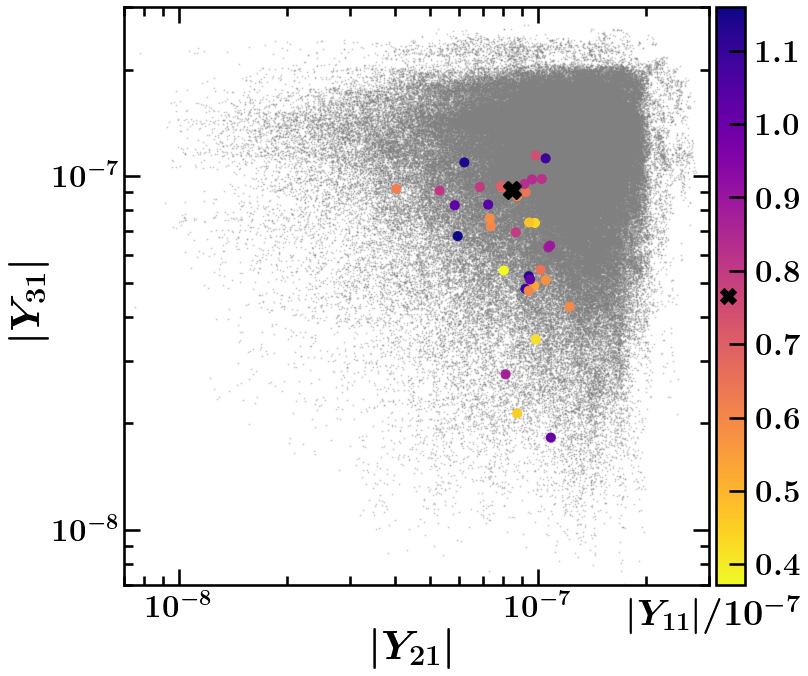}
        \caption{}
        \label{fig_all_N1}
    \end{subfigure}
    \begin{subfigure}{0.495\linewidth}
        \centering
        \includegraphics[width=1\textwidth]{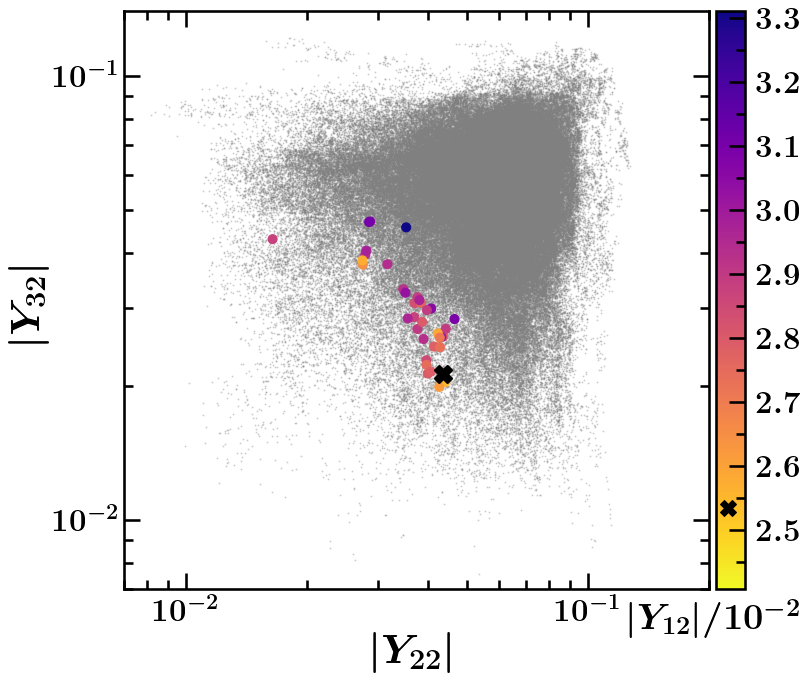}
        \caption{}
        \label{fig_all_N2}
    \end{subfigure}
    \begin{subfigure}{0.495\linewidth}
        \centering
        \includegraphics[width=1\textwidth]{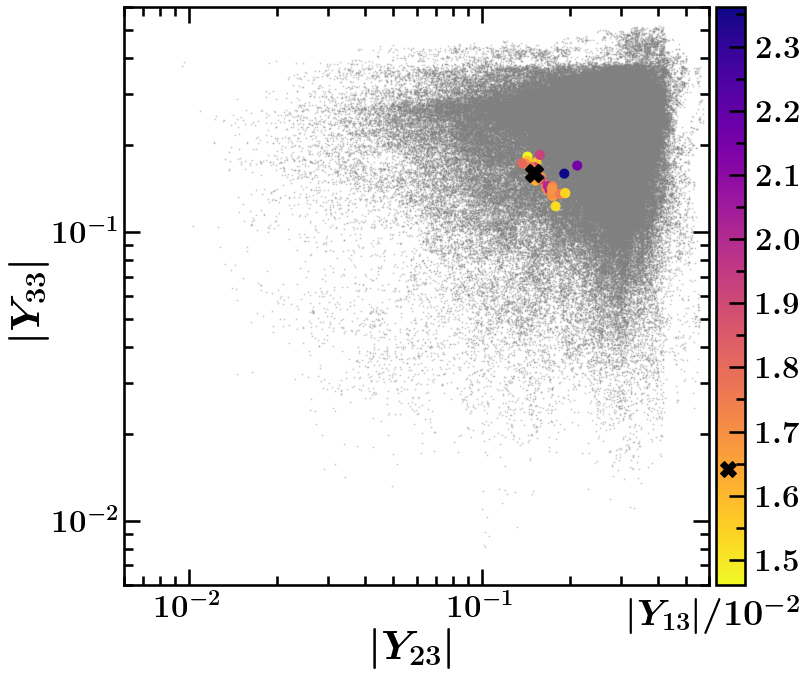}
        \caption{}
        \label{fig_all_N3}
    \end{subfigure}
    \begin{subfigure}{0.495\linewidth}
        \centering
        \includegraphics[width=1\textwidth]{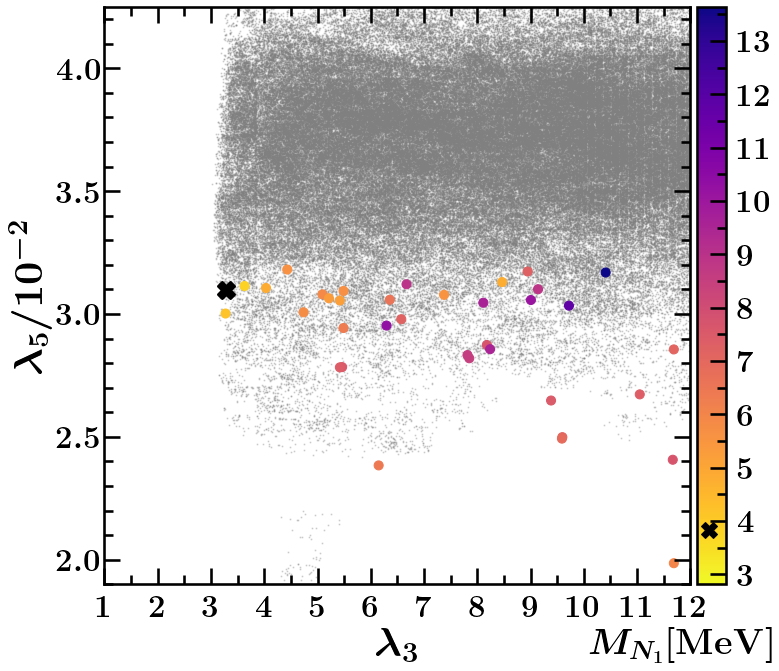}
        \caption{}
        \label{fig_all_only_l3}
    \end{subfigure}
    \caption{Gray dots represent the allowed parameter space satisfying correct baryon asymmetry and DM relic abundance while obeying the FIMP lower mass bound presented in terms of the Yukawa couplings of \eqref{fig_all_N1} $N_1$, \eqref{fig_all_N2} $N_2$, \eqref{fig_all_N3} $N_3$ RHN and \eqref{fig_all_only_l3} quartic couplings between Higgs and the inert scalar doublet. The coloured points additionally satisfies all the neutrino oscillation observables, cosmological upper bound on $\sum m_\nu$ and LFV constraints. The benchmark point given in Table \ref{tab:bp_details} is indicated by \ding{54}.}
    \label{fig_all}
\end{figure}
}


\section{Introduction} \label{sec:intro}

The origin of matter in the Universe remains a mystery within the standard paradigm of particle physics and cosmology. However, we have robust evidences to believe in the existence of two components of the gravitating matter viz. the luminous baryons that make up the stars that light up the night sky ($15\%$) and the dark matter (DM) that form the galactic halos ($85\%$) that hosts these stars \cite{Cirelli:2024ssz}. This is further compounded with the observations of matter antimatter asymmetry at the cosmological scales as evidenced by the CMB/BBN studies \cite{Planck:2018vyg, ParticleDataGroup:2024cfk}. Today we have a sound theoretical and ever increasing experimental evidence of the zoo of particles that populate the visible sector, however the particulate nature of the dark sector remains a speculation. It is possible to consider scenario where the dark sector is dominated by primordial black holes \cite{Carr:2020xqk}, MACHOs \cite{Griest:1990vu, Gates:1995js} or may reflects a modified description of gravity \cite{Milgrom:1983ca}. In this work we will presume a multipartite dark sector in similar spirit to the visible particle spectrum. A common origin of dark matter and visible matter has the mandate to explain both the origin of the relative density of the two sectors $\Omega_{DM}/\Omega_b \sim 5$ and the observed matter-antimatter asymmetry in the visible Universe $\eta_b \sim 10^{-10}$ within a natural unified theory.

Cogenesis is a proposed framework that addresses these issues by linking the process of baryogenesis with dark matter production in the early Universe \cite{Davoudiasl:2012uw, Dasgupta:2016odo, Bernal:2017zvx, Borah:2018uci, Mahanta:2019gfe, Dasgupta:2019lha, Cui:2020dly, Chu:2021qwk}. In this article, we provide a novel version of cogenesis originating from the scotogenic neutrino mass model giving rise to a multipartite dark sector. In contrast to the usual approach to consider an asymmetric freeze-in of baryons and thermal freeze-out of DM \cite{Kashiwase:2012xd, Hugle:2018qbw, Baumholzer:2018sfb, Heisig:2024mwr}, we present a mechanism where the freeze-in production of visible and a non thermal dark sector are conjoint to the same CP violating decay of a $\mathbb{Z}_2$-odd heavy sterile neutrino within the minimal scotogenic framework. The CP violation of the decay dictates the present day baryon asymmetry through leptogenesis. The dark matter is generated by subsequent decay of the dark sector particle to the lightest stable state as sketched in Fig. \ref{fig:schematic}. This decay is delayed due to the competition between the  scattering and decay processes that is crucial in determining the relic density for the freeze-in dark matter to remain consistent with cosmological constraints from large scale structure formation.

We sketch the minimal version of this framework within a scotogenic model having three $\mathbb{Z}_2$-odd right handed neutrinos (RHNs) and a single $\mathbb{Z}_2$-odd scalar doublet. With the masses of the heavier states set around $\mathcal{O}(10^9)$ GeV, we demonstrate how this framework can simultaneously generate the observed baryon asymmetry and a freeze-in dark matter in the sub-GeV mass scale that saturates the relic density while remaining consistent with the neutrino oscillation parameters \cite{Esteban:2020cvm}.
\begin{figure}[t]
\centering
\resizebox{0.475\textwidth}{!}{
\begin{tikzpicture}[every text node part/.style={align=center}]
	\tikzstyle{boxstyle} = [rectangle, rounded corners, text centered, draw = black];
	\tikzstyle{ellipsestyle} = [ellipse, text centered, draw = black, thick];
	\tikzstyle{circlestyle} = [circle, text centered, draw = black, thick];
	\node (start) [rounded rectangle, fill=green!25, minimum height=1cm, minimum width=7.5cm, draw=black, thick] {\large \bfseries Primordial bath};
	\node (N2) [ellipsestyle, below = 1.35cm of start, fill = red!10] {Primordial \\[-0.25ex] dark state ($N_2$)};
	\draw[thick, double, latex-latex, {Latex[length=2.5mm, width=3.5mm]}-{Latex[length=2.5mm, width=3.5mm]}] ($(start.south)-(0, 0.0)$) -- (N2.north) node [midway, right] () {thermal \\[-0.25ex] equilibrium};
	\node (cp) [circlestyle, below = 0.75cm of N2, fill = gray!35] {\cancel[thick]{CP}\\[-0.25ex] decay};
	\draw [thick, latex-, -{Latex[length=1.5mm, width=2mm]}] {(N2.south) -- (cp.north)};
	\node (eta) [boxstyle, thick, below left = 0.75cm and 0.75cm of cp, fill = red!10] {Next-to-lightest \\[-0.25ex] dark particle ($\eta$)};
	\draw [thick, latex-, -{Latex[length=1.5mm, width=2mm]}] {($(cp.225)+(0.005, 0.005)$) -- ($(eta.17.5)$)};
	\draw [thick, double, latex-latex, {Latex[length=2.5mm, width=3.5mm]}-{Latex[length=2.5mm, width=3.5mm]}] (eta.north) -- ($(start.south) - (2.865cm, 0)$) node[midway, above, sloped] () {scatterings} node[midway,below,sloped] () {$\eta \overline{\eta} \leftrightarrow H \overline{H}$} ;
	\node (asymL) [boxstyle, thick, below right = 0.75cm and 0.75cm of cp, fill = green!25] {Lepton \\[-0.25ex] asymmetry ($\Delta L$)};
	\draw [thick, latex-, -{Latex[length=1.5mm, width=2mm]}] {($(cp.315)+(0.005, 0.005)$) -- ($(asymL.162.5)$)};
	\draw [thick, double, latex-latex, {Latex[length=2.5mm, width=3.5mm]}-{Latex[length=2.5mm, width=3.5mm]}] (eta.east) -- (asymL.west) node [midway, above] () {Transfer of} node [midway, below] () {asymmetry};
	\node (N1) [boxstyle, thick, below = 1.15cm of eta, fill = red!10] {Stable dark \\[-0.25ex] matter ($N_1$)};
	\draw [thick, latex-, -{Latex[length=2.5mm, width=3.5mm]}] (eta.south) -- (N1.north) node [midway, right] () {delayed \\[-0.25ex] decay};;
	\node (asymB) [boxstyle, thick, below = 1.15cm of asymL, fill = green!25] {Baryon \\[-0.25ex] asymmetry ($\Delta B$)};
	\draw [thick, latex-, -{Latex[length=2.5mm, width=3.5mm]}] (asymL.south) -- (asymB.north) node [midway, left] () {sphaleron \\[-0.25ex] effects};
\end{tikzpicture}
}
\caption{Cogenesis of matter within the minimal scotogenic model}
\label{fig:schematic}
\end{figure}
Expectedly the magnitude of CP violation in the decay of the primordial dark state $(N_2)$ sets the baryon asymmetry parameter. We will demonstrate that within this framework, evolution of the abundances can be tracked across two distinct regime. One where the scattering  between the $\mathbb{Z}_2$-odd heavy species  with the primordial thermal bath dominate and then at a later epoch the decay within the multipartite dark sector dominates.  Interestingly this competition within the primordial dark sector interaction dynamically set the ratio $\Omega_{DM}/\Omega_b \sim 5$  while providing a cosmological viable sub-GeV  dark matter  that can saturate the relic density while remaining consistent with structure formation bounds. We identify the region of parameter space that is consistent with dark matter, baryon asymmetry and all the neutrino oscillation data. We will demonstrate that these points remain orders of magnitude below the present bounds from lepton flavor violating (LFV) processes.

The article is arranged as follows. We start by describing the model in section \ref{sec:model}. We next discuss the origin of cogenesis mechanism simultaneously driving both the baryon asymmetry via leptogenesis and DM in section \ref{sec:cogenesis}. We discuss the region of parameter space that is consistent with neutrino oscillation data and relevant LFV constraints in section \ref{sec:nu_and_lfv}. We finally conclude in section \ref{sec:conclusion}.



\section{Model} \label{sec:model}

\tabFieldContent

In this section we briefly review the minimal Scotogenic model \cite{Ma:2006km}.   We make minimal extension of the SM by including a $\mathbb{Z}_2$-odd sector composed of a scalar $SU(2)_L$ doublet $(\eta)$ and a set of  hierarchical  heavy sterile neutrinos $(N_i).$ The charges of all the relevant particles are summarised in Table \ref{tab:Field_Content} and we will assume $M_{N_3} > M_{N_2} \gg M_{N_1}$. The  lepton number violating Yukawa interactions consistent with the charge assignments of Table \ref{tab:Field_Content}  can be written as
\begin{equation}
    \mathcal{L}_\text{Yuk} = - Y_{\alpha i} \overline{L}_\alpha \widetilde{\eta} N_i + \text{ h.c.} \ ,
    \label{yukawa}
\end{equation}
where $\widetilde{\eta} = i \sigma_2 \eta^*$ and $Y_{\alpha i}$ are complex Yukawa couplings.  This interaction is responsible for a novel cogenesis framework where  the same  CP violating decay of the heaviest sterile neutrino $N_2 \rightarrow \eta + L$ populates both the dark and visible sector. In the presence of an inert doublet the extended scalar potential can be written as
\begin{equation} \label{eq:Scalar_Potential}
\begin{aligned}
    V(H, \eta) = &-m_H^2 H^\dagger H + \dfrac{\lambda_H}{2} (H^\dagger H)^2 + m_\eta^2 \eta^\dagger \eta + \dfrac{\lambda_\eta}{2} (\eta^\dagger \eta)^2 \\
    &+ \lambda_3 \left( \eta^\dagger \eta \right)\left( H^\dagger H \right) + \lambda_4 \left( \eta^\dagger H \right)\left( H^\dagger \eta \right) + \dfrac{1}{2} \lambda_5 \left[ \left(\eta^\dagger H \right)^2 + \text{ h.c.} \right] \ ,
\end{aligned}
\end{equation}
where in contrast to the usual SM Higgs sector, the positive mass term of the inert scalar doublet preserves  the $\mathbb{Z}_2$  symmetry by preventing the  inert doublet from obtaining a vacuum expectation value (vev).  The $\mathbb{Z}_2$-odd  state $\eta$ couples to the visible sector through the Higgs portal couplings $\lambda_{3/4/5}.$ In the scenario where $\eta$ is the lightest dark sector state we can map this to a viable weakly interacting massive particle-like Higgs portal DM and has been studied extensively in the literature \cite{Banerjee:2021anv, Banerjee:2021hal}. However in this article we will consider it to be the next-to-lightest stable particle (NLSP) in the dark sector. Within this context we will demonstrate that the  Higgs portal couplings play a crucial role in thermalising the NLSP before it decays to the DM state. These interaction  of the NLSP with the primordial soup plays a key role in producing a viable sub-GeV dark matter state. Explicitly in terms of the components the scalar fields can be written as
\begin{equation}
    H = \begin{pmatrix} \phi^+ \\ \phi^0 \end{pmatrix} \ ; \ \eta = \begin{pmatrix} \eta^+ \\ \eta^0_R + i \eta^0_I \end{pmatrix} \ ,
\end{equation}
and without any loss of generality the vacuum configuration can be  identified as
\begin{equation}
    \langle H \rangle = \dfrac{v_H}{\sqrt{2}} \ ; \langle \eta \rangle = 0 .
\end{equation}
After the electroweak symmetry breaking the masses of the charged, real and imaginary components are given as
\begin{equation}
    M_{\eta^\pm}^2 = m_\eta^2 + \lambda_3 v_H^2 \ ; \ M_{\eta_R}^2 = M_0^2 + \lambda_5 \dfrac{v_H^2}{2} \ ; \ M_{\eta_I}^2 = M_0^2 - \lambda_5 \dfrac{v_H^2}{2} \ ,
\end{equation}
where $M_0^2 = m_\eta^2 + (\lambda_3 + \lambda_4) v_H^2/2$. It is crucial to note that the mass splitting between the real and imaginary component of the neutral part of the inert doublet is directly proportional to $\lambda_5$ which plays an important role in the generation of neutrino mass radiatively.  This will eventually correlate the neutrino sector parameters with the DM relic density in this minimal framework.

\figNeutrinoMassFeynmanDiagram

The masses of the SM neutrinos are generated radiatively at one loop as shown in Fig. \ref{fig:Neutrino_Mass_Feynman_Diagram} and is given by \cite{Escribano:2020iqq}
\begin{equation} \label{eq:neutrino_mass_matrix}
\begin{aligned}
    (m_\nu)_{\alpha \beta} = \dfrac{\lambda_5 v_H^2}{32 \pi^2} &\sum_{i = 1}^3 \dfrac{Y_{\alpha i}Y_{\beta i}}{M_{N_i}} \Bigg[ \dfrac{M_{N_i}^2}{M_0^2 - M_{N_i}^2} + \left( \dfrac{M_{N_i}^2}{M_0^2 - M_{N_i}^2} \right)^2 \log \left( \dfrac{M_{N_i}^2}{M_0^2} \right) \Bigg] \, ,
\end{aligned}
\end{equation}
which is directly proportional to the non-self hermitian part of the potential given in Eq. \ref{eq:Scalar_Potential} as stated above. The solar and atmospheric neutrino mass scales are generated from the heavy seesaw masses of $N_2$ and $N_3$.  The Yukawa structure in  Eq. \ref{yukawa} has 18 parameters and are responsible to generate the neutrino oscillation parameters \cite{Esteban:2020cvm}. The lightest RHN $N_1$ which emerges as the DM candidate have a negligible contribution towards the neutrino mass due to its tiny Yukawa couplings \cite{Molinaro:2014lfa} and  will be demonstrated to be produced by late time decay of the $\eta$ state producing a cosmologically viable sub-GeV  freeze-in dark matter.


\section{Cogenesis} \label{sec:cogenesis}

\begin{figure*}[h]
	\centering
    \begin{subfigure}{0.25\linewidth}
		\centering
		\begin{tikzpicture}[baseline=(current bounding box.center)]
		\begin{feynman}
			\vertex (a) {\large $N_2$};
			\vertex [right = of a] (b);
			\vertex [above right=of b] (c) {\large $L$};
			\vertex [below right=of b] (d) {\large $\eta$};
			\diagram{
				(a) --[plain, very thick] (b) -- [fermion, very thick] (c);
				(b) --[scalar, very thick] (d);
					};
		\end{feynman}
		\end{tikzpicture}
		\caption{}
		\label{fig:N2_decay_tree}
    \end{subfigure} \hspace{-1em}
    \begin{subfigure}{0.325\linewidth}
        \centering
        \begin{tikzpicture}[baseline=(current bounding box.center)]
        \begin{feynman}
			\vertex (a) {\large $N_2$};
			\vertex [right = 1.25cm of a] (b);
			\vertex [right = of b] (c);
			\vertex [right = 1.25cm of c] (d);
			\vertex [above right=of d] (e) {\large $L$};
			\vertex [below right=of d] (f) {\large $\eta$};
			\diagram{
				(a) --[plain, very thick] (b);
				(b) --[scalar, very thick, half left, edge label = $\overline{\eta}$] (c);
				(b) --[anti fermion, very thick, half right, edge label' = $\overline{L}$] (c);
				(c) --[plain, insertion = 0.5, very thick, edge label' = $N_3$] (d) -- [fermion, very thick] (e);
				(d) --[scalar, very thick] (f);
					};
        \end{feynman}
        \end{tikzpicture}
        \caption{}
        \label{fig:N2_decay_loop_self}
    \end{subfigure} \hspace{2em}
    \begin{subfigure}{0.325\linewidth}
        \centering
        \begin{tikzpicture}[baseline=(current bounding box.center)]
		\begin{feynman}
			\vertex (a) {\large $N_2$};
			\vertex [right = of a] (b);
			\vertex [above right = 1.9cm of b] (c);
			\vertex [below right = 1.9cm of b] (d);
			\vertex [right = of c] (e) {\large $L$};
			\vertex [right = of d] (f) {\large $\eta$};
			\diagram{
				(a) -- [plain, very thick] (b) -- [scalar, very thick, edge label = { $\overline{\eta}$}] (c) -- [fermion, very thick] (e);
				(b) -- [anti fermion, very thick, edge label' = {$\overline{L}$}] (d) -- [scalar, very thick] (f);
(c) -- [insertion = 0.5, very thick, edge label = {\large $N_3$} ] (d) ;
					};
		\end{feynman}
		\end{tikzpicture}
		\caption{}
		\label{fig:N2_decay_loop_vertex}
    \end{subfigure}
    \caption{Tree level (\ref{fig:N2_decay_tree}) and the leading one loop (\ref{fig:N2_decay_loop_self} and \ref{fig:N2_decay_loop_vertex}) decay of $N_2$ involving $N_3$.}
    \label{fig:N2_decay}
\end{figure*}

\noindent
In the presence of complex Yukawa couplings, the CP asymmetry parameter is generated from the interference of tree and 1-loop level decay of the RHNs. Choosing a hierarchical spectrum for the masses of the RHNs i.e. $M_{N_3} > M_{N_2} \gg M_{N_1}$, the leptogenesis is dominated by the decay of $N_2$ to a lepton and an inert doublet as shown in Fig. \ref{fig:N2_decay} and the corresponding asymmetry parameter is given by
\begin{align} \label{eq:asym_param}
	\epsilon &= \sum_{\alpha} \dfrac{\Gamma(N_2 \rightarrow L_\alpha \eta) - \Gamma(N_2 \rightarrow \overline{L}_\alpha \overline{\eta})}{\Gamma(N_2 \rightarrow L_\alpha \eta) + \Gamma(N_2 \rightarrow \overline{L}_\alpha \overline{\eta})} \nonumber \\
	&= \dfrac{\Im \left( [Y^\dagger Y]^2_{32} \right) }{8\pi \, [Y^\dagger Y]_{22}} \sqrt{x} \left[ \dfrac{x - 2}{x - 1} - (1 + x) \log \left( 1 + \dfrac{1}{x} \right) \right] \ ,
\end{align}
where $x = M_{N_3}^2/M_{N_2}^2$. Note that the heaviest RHN ($N_3$) appearing in the loop is crucial for the generation of a sizable primordial asymmetry as the contribution coming from the lightest RHN ($N_1$) is numerically insignificant. Once these asymmetries are generated they evolve according to the collisional Boltzmann equations to produce a remnant present day lepton asymmetry eventually translating to a net baryon asymmetry via standard sphaleron processes \cite{Khlebnikov:1988sr, Harvey:1990qw}.

Interestingly, the same out of equilibrium decays of the $N_2$ produce the inert doublet $\eta$ which populates a partially asymmetric dark sector in the early Universe. The inert doublet couples with the SM Higgs  as can be seen from Eq. \ref{eq:Scalar_Potential} and thermalises with the primordial soup. Consequently, the interplay between the scattering with the Higgs and its suppressed  coupling to the  lightest dark sector state $N_1$ fixes the onset of the decay of the inert doublet to the freeze-in dark matter $N_1$.  This provides a handle to  set the relic density of a sub-GeV DM.

In order to investigate the evolution of DM and lepton asymmetry  simultaneously we systematically setup and numerically simulate the coupled system of Boltzmann equations tracking the abundances of $N_2$, $\Sigma \eta = \eta + \overline{\eta}$, $\delta \eta = \eta - \overline{\eta}$, the $B - L$ asymmetry and the $N_1$. The relevant integro-differential equations are given by
\begin{subequations} \label{eq:boltz_full}
\begin{align}
\dot{Y}_{N_2} &= - \dfrac{1}{4} \left(4\yNtwo - 2 \ySE \right) \gamNtwo - \left(\yNtwo - \yNone \right) \left(2 \gamNtwoLTONoneL + \ySE \gamNtwoETONoneE \right) \nonumber \\
{}& -\dfrac{1}{4} \sum_{i = 1}^2 (4 \yNtwo y_{N_i} + \ydL^2 - 4) \gamma^{N_2 N_i}_{L \overline{L}} - \dfrac{1}{4} \sum_{i = 1}^2 (4 \yNtwo y_{N_i} - \ySE^2 + \ydE^2) \gamma^{N_2 N_i}_{\eta \overline{\eta}} \ , \\
\dot{Y}_{\Sigma \eta} &= \dfrac{1}{4} \Big( 4\yNtwo - 2\ySE - \ydE \ydL \Big) \gamNtwo - \left( \ySE - 2 \yNone \right) \gamE \nonumber \\
{}& + \dfrac{1}{2} \sum_{j \geqslant i}^2 \sum_{i = 1}^2 \left( 4 y_{N_i} y_{N_j} - \ySE^2 + \ydE^2 \right) \gamma^{N_i N_j}_{\eta \overline{\eta}} - \dfrac{1}{2} \left( \ySE^2 + \ydE^2 - \ydL^2 - 4 \right) \gamEETOLbLb  \nonumber \\
{}& - \dfrac{1}{2} \left( \ySE^2 - \ydE^2 - 4 \right) \gamEEbTOHHb - \dfrac{1}{2} \left( \ySE^2 + \ydE^2 - 4 \right) \gamEETOHH \ , \\
\dot{Y}_{\delta \eta} &= \dfrac{\epsilon}{4} \Big( 4 \yNtwo + (\ydL -4)\ydE + 2 (\ydL - 1) \ySE \Big) \gamNtwo - (2 \ydE + \ydL \ySE) \gamLETOLbEb \nonumber \\
{}& - (2 \ydL + \ydE \ySE) \gamEETOLbLb - (\yNone \ydL + \ydE) \gamE - \ydE \ySE \gamEETOHH \ , \\
\dot{Y}_{\delta (B - L)} &= -\dfrac{1}{4} \epsilon \Big( 4 \yNtwo - 2 \ySE - \ydL \ydE \Big) \gamNtwo \nonumber \\
{}& + (\yNone \ydL + \ydE) \gamE + (2 \ydL + \ydE \ySE) \gamEETOLbLb + (2 \ydE + \ydL \ySE) \gamLETOLbEb  \ , \\
\dot{Y}_{N_1} &= (\ySE - 2 \yNone) \gamE + (\yNtwo - \yNone) \left(2 \gamNtwoLTONoneL + \ySE \gamNtwoETONoneE \right) \nonumber \\
{}& -\dfrac{1}{4} \sum_{i = 1}^2 (4 \yNtwo y_{N_i} + \ydL^2 - 4) \gamma^{N_2 N_i}_{L \overline{L}} - \dfrac{1}{4} \sum_{i = 1}^2 (4 \yNtwo y_{N_i} - \ySE^2 + \ydE^2) \gamma^{N_2 N_i}_{\eta \overline{\eta}} \ ,
\end{align}
\end{subequations}
\tabParamRange
\tabBPdetails

\noindent
where $\dot{Y} = z \, s(z) \, H(z) dY/dz$ with $z = M_{N_2}/T$. Further $y_X = Y_X/Y_X^\eq$ has been utilised for convenience where $Y_X = n_X/s$ is the comoving number density of any species $X \in \lbrace N_2, \Sigma \eta, \delta \eta, \delta(B - L), N_1 \rbrace$ and the superscript ``eq'' denotes its thermal equilibrium value. The Hubble parameter $H(z)$ and the entropy density $s(z)$ are given by
\begin{equation}
    H(z) = \sqrt{\dfrac{8 g_*}{\pi}} \dfrac{M_{N_2}^2}{M_\text{Pl}} \dfrac{1}{z^2} \, ; \quad s(z) = \dfrac{2 \pi^2}{45} g_* \left( \dfrac{M_{N_2}}{z} \right)^2 \, ,
\end{equation}
with $g_*$ being the effective number of degrees of freedom and the Plank mass is set at $M_\text{Pl} = 1.22 \times 10^{19}$ GeV. The equilibrium number densities are given by
\begin{equation}
    n_{N_2}^\text{eq} = \dfrac{1}{\pi^2} M_{N_2}^3 \dfrac{K_2(z)}{z}\, ; \quad n_{\Sigma \eta}^\text{eq} = \dfrac{2 r_\eta}{\pi^2} M_\eta^3 \dfrac{K_2(r_\eta z)}{z} \, ; \quad n_L^\text{eq} = \dfrac{2}{\pi^2} \left(\dfrac{M_{N_2}}{z} \right)^3 \, ,
\end{equation}
with $r_\eta = M_{\eta}/M_{N_2}$ and $K_i(z)$ denotes the modified Bessel function of $i$-th order. The reaction densities related to the decay of $N_2$ and $\eta$ are respectively given by
\begin{equation}
    \gamNtwo = n_{N_2}^\text{eq} \dfrac{K_1(z)}{K_2(z)} \dfrac{M_{N_2}}{8 \pi} [ Y^\dagger Y]_{2 2} \, ; \quad \gamE = n_{\Sigma \eta}^\text{eq} \dfrac{K_1(z)}{K_2(z)} \dfrac{M_{\eta}}{8 \pi} [ Y^\dagger Y]_{1 1}
\end{equation}
where the ratio of Bessel functions accounts for the dilution factor \cite{Buchmuller:2004nz} arising from the expansion of the Universe. The reaction densities for generic $2 \leftrightarrows 2$ scattering processes in the center of mass frame is denoted as $\gamma^X_Y$ and is given by
\begin{equation}
    \gamma^X_Y \equiv \gamma(X \rightarrow Y) = \dfrac{T}{64 \pi^4} \int_{s_\text{min}}^\infty ds \sqrt{s} K_1(z \sqrt{s}) \widehat{\sigma} (s)
\end{equation}
where $\widehat{\sigma}$ is the reduced cross section \cite{Luty:1992un}.

\tabBPoutput
\figBPforReacDen
\figBPforY

We choose a benchmark point given in Table \ref{tab:bp_details} to illustrate the dynamics of cogenesis in the minimal scotogenic model. The decay reaction densities of the inert scalar doublet as well as its scatterings with the Higgs are shown in Fig. \ref{fig:reac_den}. In presence of larger quartic couplings and relatively smaller decay width of $\eta$, the scattering processes tend to keep inert doublet in thermal equilibrium delaying its decay into a lepton and $N_1$. The equality of these reaction densities occurs at $z \sim 250$ denoted by a vertical dashed gray line in Fig. \ref{fig:reac_den} after which the decay reaction density takes over populating the dark sector with $N_1$. All other reaction densities are shown in Fig. \ref{fig:reac_den_all} including the $\Delta L = 2$ scattering processes which modules the asymmetry between the dark sector and the visible sector. The evolution of various comoving number densities are shown in Fig. \ref{fig:Y} with respect to the dimensionless variable $z$. As can be seen from the figure the CP violating decay of $N_2$ generates an asymmetry in the lepton sector and in the inert doublet. The lepton asymmetry gets converted into the baryon asymmetry via standard sphaleron processes yielding the observed BAU measured by Planck \cite{Planck:2018vyg} shown via the narrow horizontal shaded band with the numerical value given in Table \ref{tab:bp_details_output}. However, the produced inert doublet rapidly thermalises with the primordial plasma owing to its high scattering rate regulated by its couplings with the SM Higgs $(\lambda_3)$ delaying its decay. As the Universe expands further, the flux suppression increases and around $z \sim 250$ the decay process starts to dominate over the scattering processes populating the dark sector with stable DM candidate $N_1$. The $\Delta L = 2$ scattering processes modulates the asymmetry between the dark sector and the visible sector.

\figParamSpaceLeptoDM

In our framework, the main constraints on the minimum mass of DM particles arise from the cosmic microwave background (CMB) and big bang nucleosynthesis (BBN) observations. For DM that annihilates through \emph{s}-wave processes, energy injection from annihilation can alter the CMB spectrum, leading to a lower bound on the DM mass, typically $m_{\chi} \gtrsim \mathcal{O}(10) \, {\rm GeV}$ \cite{Leane:2018kjk}. For other scenarios, such as \emph{p}-wave annihilation, BBN provides the strongest constraint since relativistic DM during BBN can alter the primordial element abundances, requiring the DM to be heavier than the MeV scale \cite{Sabti:2019mhn}.

In our framework, DM is produced via the \emph{freeze-in} mechanism, where the coupling to the Standard Model is so feeble that it does not affect the CMB. Additionally, since DM is never in thermal equilibrium with the primordial plasma, its abundance does not impact BBN either. Therefore, for \emph{freeze-in} DM, the lower mass limit mainly comes from structure formation bounds. Observations of dwarf galaxies, which are kpc scales, imply that the DM de-Broglie wavelength must be smaller than these structures, giving $m_\chi \gtrsim 10^{-21} \, {\rm eV}$ \cite{Zimmermann:2024xvd}. If the DM is fermionic, the Pauli exclusion principle comes into play which pushes the lower bound at keV scales \cite{Tremaine:1979we, Alvey:2020xsk}. These limits are independent of the DM production mechanism and depend mainly on whether the DM is bosonic or fermionic. In the typical \emph{freeze-in} scenarios, DM is mainly produced when the temperature of the universe is around the mass scale of the mother particle. After production, DM free-streams, and if it is too light, its kinetic energy can wash out structures. As calculated in \cite{DEramo:2020gpr}, using phase space distributions and Lyman-$\alpha$ forest data, the lower mass bound for such DM is found to be larger than $\mathcal{O}(1)$ keV.

In our framework, the production of the dark matter particle $(N_1)$ differs slightly from the usual \emph{freeze-in} scenario. Both $N_2$ and the scalar $\eta$ contribute to the production of $N_1$ during cosmic evolution, with the dominant contribution arising from the decay of the complex scalar $\eta$. Importantly, $\eta$ does not decay immediately after decoupling from the plasma; instead, most of the $N_1$ population is generated through its late decay. As shown in \cite{Decant:2021mhj}, this leads to a momentum distribution of $N_1$ that deviates from the standard \emph{freeze-in} case. Since the momentum distribution governs the free-streaming length of DM and thus impacts structure formation, the corresponding constraints are modified by the decay rate of the parent particle and the bound is given by \cite{Decant:2021mhj}
\begin{equation}\label{eq:dm_limit}
    M_{N_1} \gtrsim m_{\rm lim} \times \left( \dfrac{106.75}{g_{*S}(T_{\rm decay})} \right)^{1/3} \times (R_\eta)^{-1/2} ,
\end{equation}
where the dimensionless parameter $R_\eta = (M_{\rm Pl, red} \, \Gamma_{\eta})/M_\eta^2$, with $M_{\rm Pl, red}$ denoting the reduced Planck mass, $\Gamma_{\eta}$ is the decay width of $\eta$ and $g_{*S}$ is the effective entropy degrees of freedom at the relevant decay temperature ($T_\text{decay}$). In Eq. \ref{eq:dm_limit}, $m_{\rm lim}$ corresponds to the keV-scale lower bounds from Lyman-$\alpha$ and $\Delta N_{\rm eff}$ probes, as listed in Table 1 of \cite{Decant:2021mhj}. We adopt the most stringent value, $m_{\rm lim} = 3.9$ keV. With a typical scattering cross-section with SM leptons below $\mathcal{O}(10^{-50}) \, \text{cm}^2$ they easily remain below the present and future sensitivity of direct detection experiments \cite{XENON:2019gfn}.

In our framework, the dark matter is the lightest RHN ($N_1$) produced from the delayed decay of the inert doublet. The mass of DM is set by considering that it accounts for the entire relic abundance of DM i.e $\Omega_\text{DM} h^2 = 0.12$ \cite{Planck:2018vyg} and is found to be $M_{N_1} \simeq 3.8$ MeV which is larger than the lower mass bound on feebly interacting massive particles (FIMP) mass which is calculated to be $\simeq 3.5$ MeV according to Eq. \ref{eq:dm_limit} as explicitly mentioned in Table \ref{tab:bp_details_output}. To explore the parameter region for successful cogenesis while remaining consistent with the structure formation bounds, we perform an extensive numerical scan utilizing a dedicated Markhov Chain Monte Carlo algorithm following \cite{Sarazin:2021nwo} over the 18 free Yukawa couplings and two quartic couplings as given in Table \ref{tab:range_free_param}. The ranges of the quartic couplings and the Yukawas lie within the theoretical upper bounds obtained from tree-level unitarity \cite{Kanemura:1999xf, Akeroyd:2000wc, Hambye:2009pw, Arhrib:2012ia} and are given by $|\lambda_j| \leq 8 \pi, Y_{\alpha i} \leq \sqrt{4 \pi} \, ,$ where $\lambda_j \in \left\lbrace \lambda_3, \lambda_5 \right\rbrace$ and $\alpha, i = 1, 2$ and 3. We fix the mass scales as $M_3 = 5 \times 10^{10}$ GeV, $M_2 = 5 \times 10^9$ GeV, $M_\eta = 5 \times 10^8$ GeV along with the quartic coupling $\lambda_4 = 0$, while the other allowed parameter regions is shown in Fig. \ref{fig_lepto_only}, where the benchmark point is denoted by \ding{54}. The present and forthcoming collider experiments are not expected to provide a direct probe for the heavy states within this framework. An additional benchmark point with higher values of $\lambda_3$ is presented in Appendix \ref{app} illustrating the dynamics of even larger thermal scattering rate compared to the decay of the inert scalar doublet yielding higher DM mass while remaining safe from the structure formation bounds.


\section{Neutrino mass and LFV constraints} \label{sec:nu_and_lfv}

The neutrino mass in the scotogenic model is generated radiatively at one loop as shown in Fig. \ref{fig:Neutrino_Mass_Feynman_Diagram} and explicitly mentioned in Eq. \ref{eq:neutrino_mass_matrix}. To extract the neutrino oscillation observables one can resort to the Casas-Ibarra parametrization \cite{Casas:2001sr} for the neutrino mass matrix. Alternatively, we follow the algorithm prescribed in \cite{Adhikary:2013bma} to find out all the oscillation parameters by constructing the matrix $h = m_\nu m_\nu^\dagger$ which can be diagonalized by a unitary matrix in the following manner
\begin{equation}
    U^\dagger h U = \text{diag} \left(m_1^2, \ m_2^2, \ m_3^2 \right)
\end{equation}
where $m_1^2, m_2^2$ and $m_3^2$ are the squared eigenvalues of the neutrino mass matrix. The mixing angles and the CP phase appearing in the unitary matrix can be parametrized using the PDF convention \cite{ParticleDataGroup:2024cfk} as
\begin{equation}
    U = \begin{pmatrix}
            1 & 0 & 0 \\
            0 & c_{23} & s_{23} \\
            0 & -s_{23} & c_{23}
        \end{pmatrix} \begin{pmatrix}
                          c_{13} & 0 & s_{13} e^{-i \delta_\text{CP}} \\
                          0 & 1 & 0 \\
                          -s_{13} e^{-i \delta_\text{CP}} & 0 & c_{13}
                      \end{pmatrix} \begin{pmatrix}
                                        c_{12} & s_{12} & 0 \\
                                        -s_{12} & c_{12} & 0 \\
                                        0 & 0 & 1
                                    \end{pmatrix}
\end{equation}
where $s_{i j} \equiv \sin \theta_{i j}$ and $c_{i j} \equiv \cos \theta_{i j}$. We have chosen the normal hierarchy i.e. $m_3 > m_2 > m_1$ to find out the angles and the CP phase while remaining consistent within the $3\sigma$ interval of global fit of the experimental neutrino oscillation data \cite{Esteban:2020cvm}, cosmological upper bound on sum over neutrino masses i.e $\sum m_\nu < 0.12$ eV \cite{Planck:2018vyg} and the stringent constraint on effective neutrino mass parameter from KamLAND-Zen Collaboration i.e $m_{\beta \beta} < (0.061 - 0.165)$ eV \cite{KamLAND-Zen:2016pfg}. The benchmark point given in Table \ref{tab:bp_details} satisfies all the neutrino oscillation parameters along with the above mentioned constraints as can be seen from the numerical values quoted in Table \ref{tab:bp_details_output}.

\figMDMvsSummnu

Within the minimal scotogenic model, complex Yukawa couplings can induce processes with lepton flavor violation (LFV) such as $l_\alpha \to l_\beta \gamma$ and $l_\alpha \to 3 l_\beta$ can occur through one-loop diagrams. Additionally, complex Yukawa couplings can also generate non-zero electric dipole moments (EDMs) of charged leptons via two-loop diagrams. Along with neutrino oscillation constraints, we also implement the experimental constraints arising from such LFV processes. The branching ratio of $l_\alpha \to l_\beta \gamma$ is given by \cite{Toma:2013zsa}
\begin{equation}\label{eq:l_alpha_to_l_beta_gamma}
    {\rm Br}\left( l_\alpha \to l_\beta \gamma \right) = \dfrac{3 (4 \pi)^3 \alpha_{\rm em}}{4 G_F^2} \left| A_D \right|^2 {\rm Br} \left( l_\alpha \to l_\beta \nu_\alpha \bar{\nu_\beta} \right) ,
\end{equation}
where $\alpha_{\rm em} = e^2/4 \pi$ and $G_F$ is the Fermi constant. In Eq. \ref{eq:l_alpha_to_l_beta_gamma}, for conservative purpose, we have considered ${\rm Br} \left( l_\alpha \to l_\beta \nu_\alpha \bar{\nu_\beta} \right) = 1$ and the model-dependent 1-loop dipole contributions are encoded in $A_D$ which is given by
\begin{equation}\label{eq:A_D}
    A_D = \sum_{i=1}^3 \dfrac{ Y_{\beta i}^* Y_{\alpha i}}{2 (4 \pi)^2} \dfrac{1}{M_{\eta^+}^2} F_2(\xi_i) \, ,
\end{equation}
where $\xi_i = M_{N_i}^2 / M_{\eta^+}^2$ and the loop function $F_2$ is given by
\begin{equation}\label{eq:F_2}
    F_2 (x) = \dfrac{1 - 6 x + 3 x^2 + 2 x^3 - 6 x^2 {\rm log} x}{6 (1-x)^2} \, .
\end{equation}
The most stringent constraint on such processes arises from the $\mu \to e \gamma$ decay, with the MEG II experiment placing an upper limit on the branching ratio of ${\rm Br}(\mu \to e \gamma) < 1.5 \times 10^{-13}$ \cite{MEGII:2025gzr}. For $l_\alpha \to 3 l_\beta$ processes, both the 1-loop box and penguin diagrams can contribute, however, we have not considered penguin diagrams involving the Higgs boson, since the corresponding Yukawa couplings are suppressed. Although this suppression does not apply to processes involving $\tau$-leptons, the experimental limits on LFV in the $\tau$ sector are comparatively weaker, making this approximation valid for our analysis. The branching ratio for the $l_\alpha \to 3 l_\beta$ process can be expressed as \cite{Toma:2013zsa}
\begin{equation}
    \begin{aligned}
        {\rm Br} &\left( l_\alpha \to 3 l_\beta \right) = \dfrac{3 (4 \pi)^2 \alpha_{\rm em}^2}{8 G_F^2} {\rm Br} \left( l_\alpha \to l_\beta \nu_\alpha \bar{\nu_\beta} \right) \bigg[ \left| A_{ND} \right|^2 + \left| A_D \right|^2 \left\{ \dfrac{16}{3} \ {\rm log} \left( \dfrac{m_\alpha}{m_\beta} \right) - \dfrac{22}{3} \right\} \\
        &+ \dfrac{\left| B \right|^2}{6} + \dfrac{1}{3} \left( 2 \left| F_{RR} \right|^2 + \left| F_{RL} \right|^2 \right) + \left\{ -2 A_{ND} A_D^* + \dfrac{1}{3} A_{ND} B^* - \dfrac{2}{3} A_D B^* + {\rm h.c.} \right\} \bigg] \, ,
    \end{aligned}
\end{equation}
where the non-dipole contributions are encapsulated in $A_{ND}$ and is given by
\begin{equation}\label{eq:A_ND}
    A_{ND} = \sum_{i=1}^3 \dfrac{Y_{i \beta }^* Y_{i \alpha}}{6 (4 \pi)^2} \dfrac{1}{M_{\eta^+}^2} G_2(\xi_i) \, ,
\end{equation}
with the loop function $G_2$ given by
\begin{equation}\label{eq:G_2}
    G_2(x) = \dfrac{2 - 9x + 18 x^2 - 11 x^3 + 6 x^3 {\rm log} x}{6 (1 - x)^4} \, .
\end{equation}
The factors $F_{RR}$ and $F_{RL}$ are given by
\begin{equation}
    F_{RR} = F \dfrac{g_R}{g_2^2 \sin^2 \theta_W M_Z^2} \, ; \quad F_{RL} = F \dfrac{g_L}{g_2^2 \sin^2 \theta_W M_Z^2} \, ,
\end{equation}
where $g_{L(R)}$ corresponds the $Z$-boson couplings to the left (right) charged leptons with $g_2$ being the weak coupling constant and $\theta_W$ is the Weinberg angle. The factor $F$ is given by
\begin{equation}\label{F}
    F = \sum_{i=1}^3 \dfrac{Y_{\beta i}^* Y_{\alpha i}}{2 (4 \pi)^2} \dfrac{m_{l_\alpha} m_{l_\beta}}{M_{\eta^{+}}^2} \dfrac{g_2}{\cos \theta_W} F_2(\xi_i) \, ,
\end{equation}
and the form of the coefficient $B$ is taken from \cite{Toma:2013zsa}. The strongest constraint on these decays comes from the process $\mu \to 3 e$, with the current limit from the SINDRUM experiment being ${\rm Br} (\mu \to 3 e) < 10^{-12}$ \cite{SINDRUM:1987nra}. The Mu3e experiment is expected to improve this sensitivity to ${\rm Br} (\mu \to 3 e) < 10^{-16}$ in the future \cite{Amarinei:2025ntv}. In our analysis, we have required the branching ratio to remain below the projected Mu3e sensitivity, which automatically satisfies the present experimental bound.

Along with the LFV processes, the CP violating Yukawa couplings present in our framework can introduce EDMs to charged leptons, which can be written as \cite{Abada:2018zra}
\begin{equation}\label{eq:EDM}
    d_\alpha = - \dfrac{e }{(4 \pi)^4 \, M_{\eta^+}^2} \sum_\beta \sum_{i,j=1}^2 \left[ J^M_{ij \alpha \beta} \, \sqrt{\xi_i \xi_i} \, I_M(\xi_i, \xi_j) + J^D_{i j \alpha \beta} \, I_D(\xi_i, \xi_j) \right],
\end{equation}
where the Majorana and Dirac quartic invariants are denoted respectively as $J^M_{ij \alpha \beta} = \mathrm{Im}\left[Y_{j \alpha}^* \, Y_{j \beta}^* \, Y_{i \beta} \, Y_{i \alpha} \right]$ and $J^D_{ij \alpha \beta} = \mathrm{Im} \left[Y_{j \alpha}^* \, Y_{j \beta} \, Y_{i \beta}^* \, Y_{i \alpha} \right]$ with the 1-loop functions denoted by $I_M(\xi_i, \xi_j)$ and $I_D(\xi_i, \xi_j)$. The upper limit on the electron EDM is at least ten orders of magnitude more stringent than for other charged leptons, with the latest bound given by the ACME Collaboration as $\left| d_e \right| < 1.1 \times 10^{-29} \, e \, {\rm cm}$ \cite{ACME:2018yjb}.

Due to the very high scale of the inert scalar doublet, the contribution to the LFV processes along with electron EDM is minuscule as can be inferred from the representative values given in Table \ref{tab:bp_details_output} for the benchmark point shown in Table \ref{tab:bp_details}. So the entire parameter space remains unconstrained after invoking the upper limits from measurements of LFV processes.

\figParamSpaceAll

Apart from these laboratory limits, there are constraints on the total mass of all the neutrino species ($\sum m_\nu$) from Planck \cite{Planck:2018vyg} as can be seen from Fig. \ref{fig_MN1_vs_summnu} where the allowed points are shown in colour. However, satisfying all the neutrino oscillation parameters along with leptogenesis and relic density while remaining consistent with the structure formation bounds significantly shrinks the allowed parameter region as is shown in Fig. \ref{fig_all}. Additionally, it is evident from the appearance of gray points in Fig. \ref{fig_all_only_l3} that the parameter region with $\lambda_3 \lesssim 3$ is highly constrained due to the stringent structure formation bound on the mass of freeze-in DM. Relaxing the mass scales of the heavy dark sector particles might open up new regions of parameter space but this leads to nothing new insights on the cogenesis framework for which we prefer to fix the mass scales.


\section{Conclusions} \label{sec:conclusion}

Cogenesis  provides a minimal framework to simultaneously address the generation of matter density of visible and dark sector in the early Universe. We explore the possibility of the minimal scotogenic neutrino mass model to drive such cogenesis from a primordial dark sector. Within this framework, for the  first time, we study the viability of  a common CP violating decay of a heavy $\mathbb{Z}_2$-odd state populating both the visible and dark sector. Thus providing a handle to address the three issues of neutrino masses, dark matter and baryon asymmetry within a simple framework.

We demonstrate that the late time decay of the next-to-lightest  dark particle  arising from an  underlying competition between decay and scattering  with the primordial soup, is crucial to evade cosmological constraints within this cogenesis framework. The boosted production of the sub-GeV dark matter from a heavy primordial dark sector state imposes strong constraints on the dark matter mass from structure formation bounds. We perform an extensive multidimensional numerical scan to identify the region of parameter space within the simple scotogenic model that is consistent with baryon asymmetry,  neutrino oscillation observables and flavour constraints while providing a viable freeze-in dark matter that saturates the relic density with sub-GeV mass scales.


\acknowledgments

We thank Deep Ghosh and Tarak Nath Maity for useful discussions. T. S. R. and R. P. acknowledge the Department of Science and Technology, Government of India, under the ANRF Grant Agreement No. MTR/2023/000469 (MATRICS) for financial assistance.

\begin{appendix}

\renewcommand{\thesection}{\Alph{section}}
\renewcommand{\theequation}{\thesection-\arabic{equation}}

\setcounter{equation}{0}

\section{Benchmark point with higher value of $\lambda_3$} \label{app}

\tabBPdetailsOld
\begin{figure}[h]
    \centering
    \begin{subfigure}{0.475\textwidth}
        \centering
        \includegraphics[width=1\linewidth]{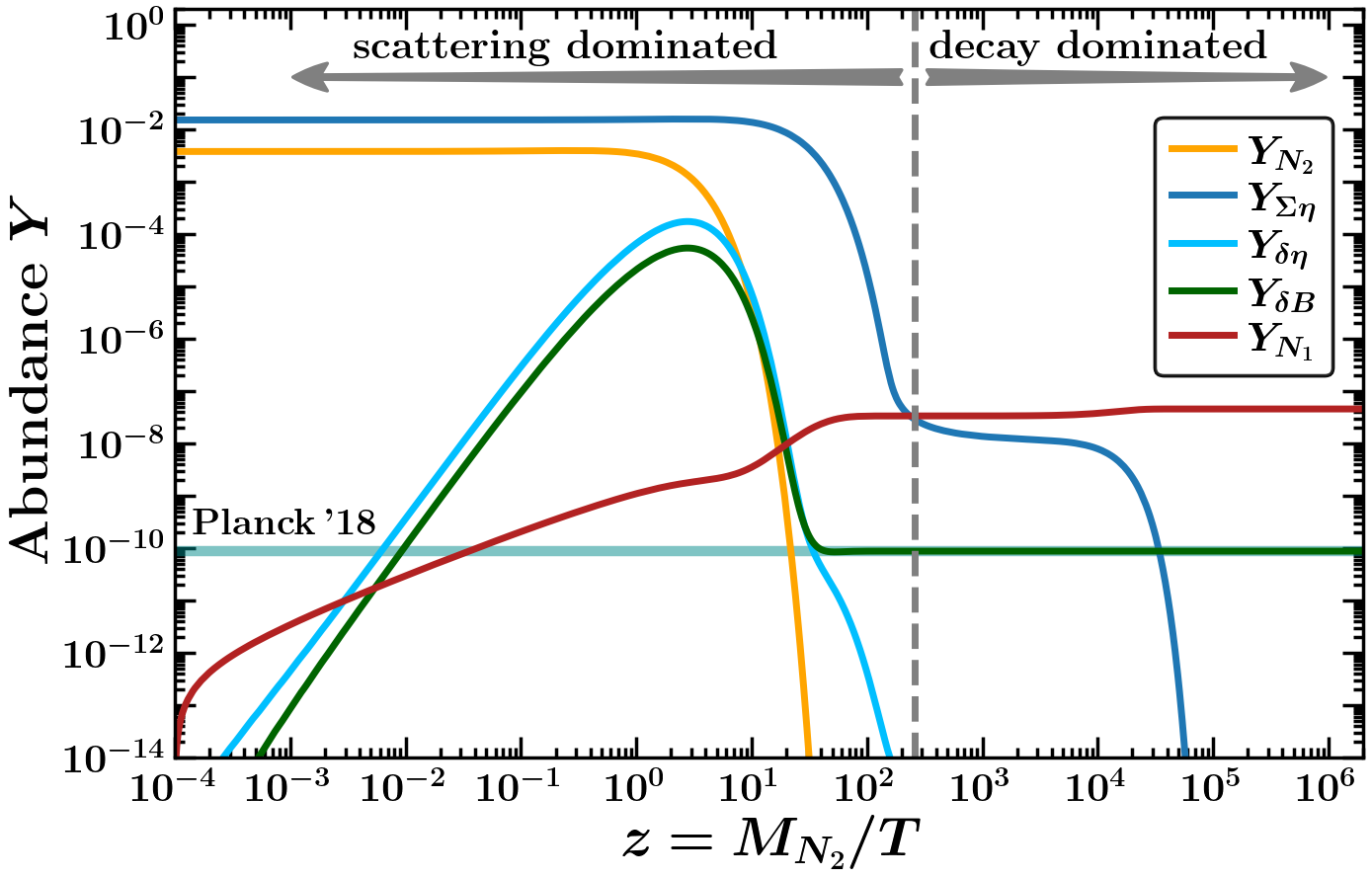}
        \caption{Same as Fig. \ref{fig:Y}.}
        \label{fig:Y_old}
    \end{subfigure}
    \begin{subfigure}{0.475\textwidth}
        \centering
        \includegraphics[width=1\linewidth]{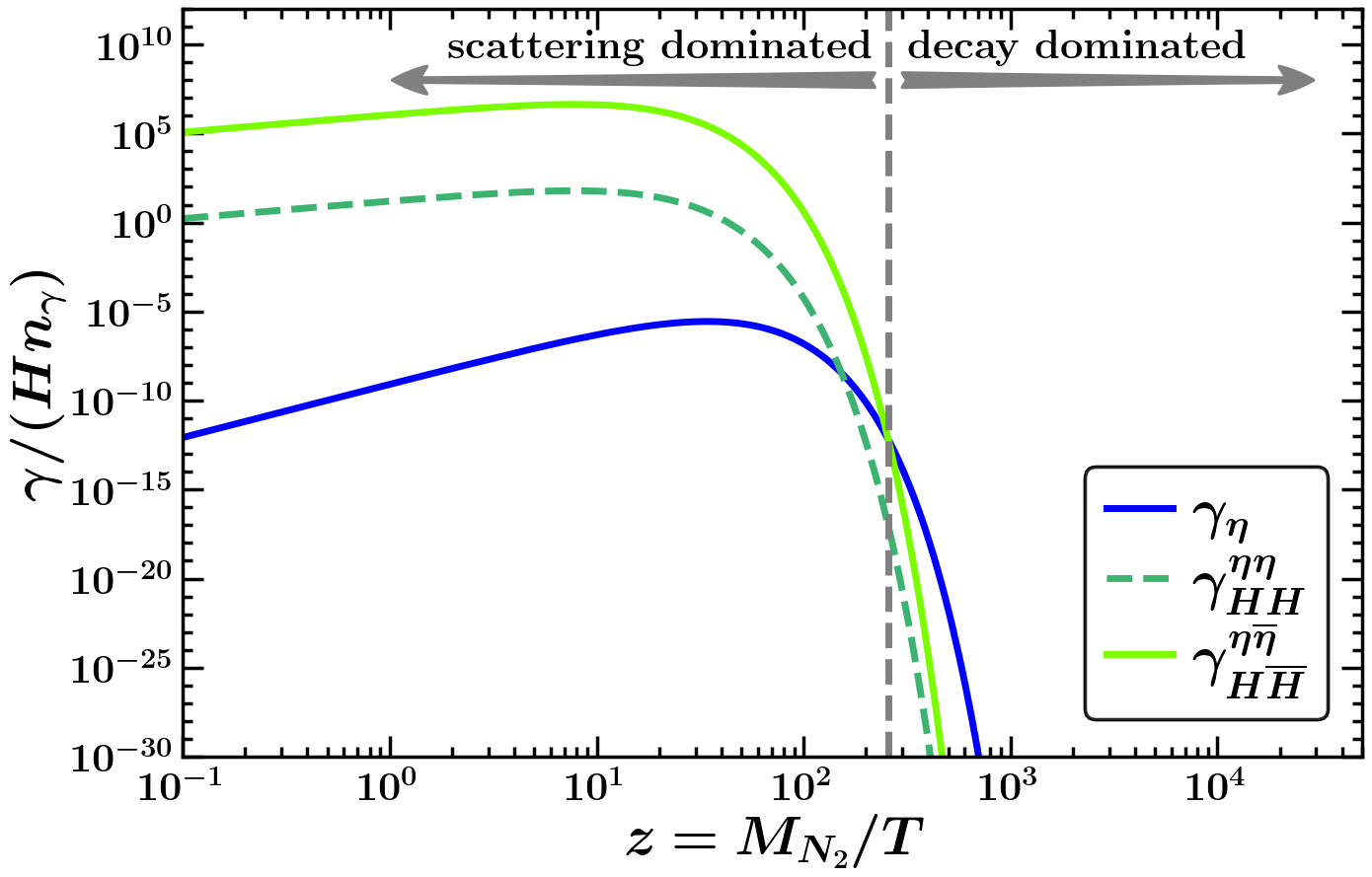}
        \caption{Same as Fig. \ref{fig:reac_den}.}
        \label{fig:reac_den_old}
    \end{subfigure}
    \caption{}
\end{figure}

\noindent
In this appendix, we present an additional benchmark point with higher values of $\lambda_3$. The mass scales of the heavy neutrinos and the inert doublet are kept same as the benchmark presented in Table \ref{tab:bp_details} with $\lambda_3 = 8.114$, $\lambda_5 = 3.046 \times 10^{-2}$ and the Yukawa couplings shown in Table \ref{tab:bp_details_old}. The evolution of various abundances and the relevant reaction densities are given in Figs. \ref{fig:Y_old} and \ref{fig:reac_den_old} respectively. Due to the increased thermal scatterings with the primordial bath, the decay of the inert scalar doublet is delayed resulting in a higher DM mass of $M_{N_1} = 9.6$ MeV, while remaining consistent with the structure formation bounds along with other constraints discussed in the main text.

\end{appendix}

\bibliographystyle{JHEP}
\bibliography{Scoto_Ma_N2_decay}

\end{document}